\documentclass[showpacs,amssymb,aps]{revtex4}
\usepackage{graphicx}

\begin{document}

\title{Vanishing magnetic mass in QED$_{3}$ with a Chern-Simons term}

\author{Ashok Das}
\author{Silvana Perez}\altaffiliation{Permanent address: Departamento
de  f\'{\i}sica, Universidade
Federal do Par\'{a}, 66075-110 Bel\'{e}m, Brasil}
\affiliation{Department of Physics and Astronomy,
University of Rochester,
Rochester, NY 14627-0171, USA}

\bigskip

\begin{abstract}
We show that, at one loop, the magnetic mass vanishes at finite
temperature  in QED in any dimension. In QED$_{3}$, even the zero
temperature part can be regularized to zero. We calculate the two loop
contributions to the magnetic mass in QED$_{3}$ with a Chern-Simons
term and show that it vanishes. We give a simple proof which shows
that the magnetic mass vanishes to all orders at finite temperature in
this theory. This proof also holds for QED in any dimension.
\end{abstract}


\maketitle

\section{Introduction}

In an earlier letter \cite{van}, we had studied the question of the
screening  mass in
the $2+1$ dimensional Abelian Higgs model with a Chern-Simons term as
well as in QED$_{3}$ with a Chern-Simons term. We had shown there that,
at one loop, the magnetic mass in QED$_{3}$ vanishes. This is quite
surprising considering the fact that the Chern-Simons term
has associated with it various magnetic phenomena \cite{deser,dunne}
and yet  the magnetic mass vanishes. In that letter \cite{van},  we had
formally  argued, based on Ward identity as well as the assumption of
analyticity of the amplitudes, that this result holds to all
orders. However, as is well known, amplitudes cease to be analytic
and infrared divergences, in general, become severe at finite
temperature \cite{das}, both of
which can invalidate a formal argument. Therefore, in this paper, we
study this question systematically in QED$_{3}$ with a Chern-Simons
term at finite temperature and give an alternate proof that the
magnetic mass indeed vanishes  to all orders. It has already been
noted \cite{brandt} that, at two loops, the parity violating part of the gauge
self-energy (correction to the Chern-Simons term) develops an infrared
divergence at finite temperature in the absence of a tree level
Chern-Simons term. It is for this reason that we study QED$_{3}$ with
a Chern-Simons term. However, as the one loop result shows \cite{van},
even  with
a tree level Chern-Simons term, the photon propagator develops a
massless pole and, therefore, the question of infrared divergence has
to be analyzed carefully.  We
note that the vanishing of the magnetic mass at finite temperature has
already  been studied
in  QED$_{4}$ \cite{fradkin}. However, infrared divergences become
more  severe as we
go to lower dimensions. The two dimensional theory (Schwinger model)
is known to be well behaved \cite{das1} and, therefore, the $2+1$ dimensional
theory is the most interesting theory to study from this point of
view. 

Our results are organized as follows. In section {\bf II}, we show
that, at one loop, the magnetic mass vanishes in QED in any dimension at
finite temperature. The additional feature of the $2+1$ dimensional
theory is that even the zero temperature contribution to the magnetic
mass can be regularized to zero. We also explicitly show that, at two
loops, QED$_{3}$ with a Chern-Simons term has vanishing contribution
to the  magnetic mass and
that there is no infrared divergence present in this amplitude at this
order even in
the absence of a tree level Chern-Simons term. In
section {\bf III}, we prove that the vanishing of the magnetic mass
holds  to all orders. In
section {\bf IV}, we present a brief summary of our results.  

\section{Explicit calculations}

Let us consider QED$_{3}$ with a Chern-Simons term described by the
Lagrangian density
\begin{equation}
L = - \frac{1}{4} F_{\mu\nu}F^{\mu\nu} + \frac{\kappa}{2}
\epsilon^{\mu\nu\lambda} A_{\mu}\partial_{\nu}A_{\lambda} +
\overline{\psi} (i\not{\!\!D} - m)\psi\label{lagrangian}
\end{equation}
where $\kappa$ is known as the Chern-Simons coefficient and the
covariant derivative is defined to be
\begin{equation}
D_{\mu}\psi = (\partial_{\mu} - ie A_{\mu})\psi
\end{equation}
In $2+1$ dimensions, the Chern-Simons term as well as the mass term
for the fermion break discrete symmetries such as parity and time
reversal \cite{deser,dunne} and, therefore, are intimately
connected. Namely, even if
there is no Chern-Simons term present at the tree level, it is
generated through radiative corrections in a massive fermion
theory \cite{redlich}. Let us note, however, that both these terms are
invariant
under charge conjugation under which
\begin{equation}
C A_{\mu} C^{-1} = - A_{\mu},\qquad C \psi C^{-1} = - \gamma^{2}
\overline{\psi}^{T}\label{chargeconjugation}
\end{equation}
As a result, the Lagrangian density in (\ref{lagrangian}) is invariant
under charge conjugation and it is the charge conjugation invariance
which, for example, makes the amplitudes with an odd
number  of photons to vanish (Furry's theorem) in this theory. We
would like to study the question of the  magnetic mass in this theory
at finite temperature.

Throughout this paper, we will use the imaginary time formalism
\cite{das,kapusta,lebellac} to
study the finite temperature effects in this theory. Therefore, we
will  consider the theory in the Euclidean
space. In such  a theory, the self-energy of the photon is independent
of the gauge fixing parameter and, in a covariant gauge, can be
parameterized to all orders as \cite{van}
\begin{equation}
\Pi_{\mu\nu} = P_{\mu\nu} \Pi_{\rm T} + Q_{\mu\nu} \Pi_{\rm L} +
\epsilon_{\mu\nu\lambda} p_{\lambda} \Pi_{\rm odd}
\end{equation}
where
\begin{equation}
P_{\mu\nu} = \tilde{\delta}_{\mu\nu} -
\frac{\tilde{p}_{\mu}\tilde{p}_{\nu}}{\tilde{p}^{2}},\qquad Q_{\mu\nu}
=
\frac{p^{2}}{\tilde{p}^{2}}\overline{u}_{\mu}\overline{u}_{\nu}\label{tensors}
\end{equation}
with
\begin{equation}
\tilde{\delta}_{\mu\nu} = \delta_{\mu\nu} - u_{\mu}u_{\nu},\qquad
\tilde{p}_{\mu} = p_{\mu} - (u\cdot p) u_{\mu},\qquad
\overline{u}_{\mu} = u_{\mu} - \frac{u\cdot p}{p^{2}} p_{\mu}
\end{equation}
Here $u_{\mu}$ represents the velocity of the heat bath which, in the
rest frame, takes the form $u_{\mu} = (1,0,0)$.

There are several things to note here. First, $\Pi_{\rm T}$ and
$\Pi_{\rm L}$, lead respectively to the transverse and the
longitudinal masses for the photon. While the longitudinal mass is
responsible for the screening of charges, it is the  transverse mass which is
related to the magnetic mass of the photon. Second, the presence of
the parity odd term, in the self-energy, is a consequence of the fact
that parity is violated in this theory. From the form of the tensors in
(\ref{tensors}), it is easy to see that we can write, in general, the
magnetic mass for the photon (which is defined in the static limit) as
\begin{equation}
\Pi_{\rm T} (0) = \frac{1}{D-2}\,\delta_{ij} \Pi_{ij} (0) =
\frac{1}{D-2}\, \delta_{ij} \Pi_{ij}^{\rm PC} (0),\qquad D >
2\label{definition} 
\end{equation}
where $D$ represents the number of space-time dimensions. Namely, the
magnetic mass is determined completely from the parity conserving part
of the self-energy and does not depend on the parity violating
structure. 

We note that in the imaginary time formalism, the tree level fermion propagator
has the form
\begin{equation}
S^{(0)} (p) = \frac{1}{\not{\!p} + m} =  \frac{- \not{\!p} + m}{p^{2} +
m^{2}}\label{fermionpropagator}
\end{equation}
where, in the Euclidean space, we work with
\begin{equation}
\gamma_{0} = i \sigma_{3},\qquad \gamma_{1} = i \sigma_{1},\qquad
\gamma_{2} = i \sigma_{2}
\end{equation}
and 
\begin{equation}
p_{0} = (2n + 1) \pi T = \frac{(2n+1)\pi}{\beta}
\end{equation} 
\begin{figure}
\centerline{\includegraphics[width = 14cm, height = 9cm]{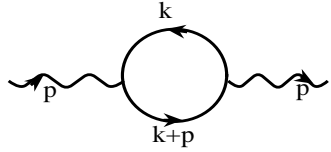}}
\vskip -4cm
\caption{One loop diagram for photon self-energy}
\end{figure}

With these, we note that the photon self-energy in QED at one loop, in an
arbitrary dimension, takes the form (see fig. 1)
\begin{eqnarray}
\Pi_{\mu\nu}^{(1)} (p) & = & e^{2} \int \frac{d^{D}k}{(2\pi)^{D}}\,{\rm
tr}\,\gamma_{\mu} S^{(0)} (k+p) \gamma_{\nu} S^{(0)} (k)\nonumber\\
 & = & \frac{e^{2}}{\beta} \sum_{n} \int
\frac{d^{D-1}k}{(2\pi)^{D-1}}\, {\rm tr} \gamma_{\mu} S^{(0)} (k+p)
\gamma_{\nu} S^{(0)} (k)
\end{eqnarray}
In $D$-dimensional Euclidean space, the trace over the gamma matrices
takes the form 
\begin{equation}
{\rm tr} \gamma_{\mu}\gamma_{\nu} = -
2^{[\frac{D}{2}]}\,\delta_{\mu\nu} = - C(D)\,\delta_{\mu\nu}
\end{equation}
where $[\frac{D}{2}]$ represents the floor of
$\frac{D}{2}$. Evaluating the Dirac trace and performing the sum over
the discrete frequencies, we obtain, in the static limit
\begin{eqnarray}
\Pi_{\rm T}^{(1)} (0) & = & \delta_{ij} \Pi_{ij}^{(1)} (0) = -
\frac{C(D)e^{2}}{2} 
\int \frac{d^{D-1}k}{(2\pi)^{D-1}}\left((D-1) +
\frac{\vec{k}^{2}}{\omega_{k}}\frac{\partial}{\partial
\omega_{k}}\right)
\left(\frac{\tanh (\frac{\beta\omega_{k}}{2})}{\omega_{k}}\right)\nonumber\\
 & = & - \frac{C(D)e^{2}}{2^{D-1} \pi^{\frac{D-1}{2}}
\Gamma(\frac{D-1}{2})} \int_{0}^{\infty} dk\,k^{D-2} \left((D-1)+
\frac{k^{2}}{\omega_{k}} \frac{\partial}{\partial \omega_{k}}\right)
\left(\frac{\tanh
(\frac{\beta\omega_{k}}{2})}{\omega_{k}}\right)\label{full}
\end{eqnarray}
where we have defined $k = |\vec{k}|$ and $\omega_{k} = \sqrt{k^{2}
+ m^{2}}$. We can now identify the temperature dependent part of
(\ref{full})  to be
\begin{equation}
\Pi_{\rm T}^{(1)(\beta)} (0) = \frac{C(D)e^{2}}{2^{D-2} \pi^{\frac{D-1}{2}}
\Gamma(\frac{D-1}{2})} \int_{0}^{\infty} dk\,k^{D-2} \left((D-1)+
\frac{k^{2}}{\omega_{k}} \frac{\partial}{\partial \omega_{k}}\right)
\left(\frac{n_{F}(\omega_{k})}{\omega_{k}}\right)\label{temp}
\end{equation}
where 
\begin{equation}
n_{F}(\omega_{k}) = \frac{1}{e^{\beta \omega_{k}} + 1}
\end{equation}
is the Fermi-Dirac distribution function. The expression on the right
hand side of (\ref{temp}) is easily seen to vanish, namely,
\begin{equation}
\Pi_{\rm T}^{(1)(\beta)} (0) = \frac{C(D)e^{2}}{2^{D-2} \pi^{\frac{D-1}{2}}
\Gamma(\frac{D-1}{2})} \int_{m}^{\infty}
d\omega_{k}\,\frac{\partial}{\partial
\omega_{k}}\left((\omega_{k}^{2}-m^{2})^{\frac{D-1}{2}}\,
\frac{n_{F}(\omega_{k})}{\omega_{k}}\right) = 0,\qquad D\geq 2
\end{equation}

This shows that, at finite temperature, the one loop correction to the
magnetic mass vanishes in any dimension $D\geq 2$. We note that, at
one loop, there is no infrared divergence since we are considering a
massive fermion. In general, the zero  temperature part in
(\ref{full}) has an ultraviolet  divergence. However, in
$2+1$  dimensions,
there is the added interesting feature that the zero temperature part
can be regularized to zero within the framework of Pauli-Villars
regularization or dimensional regularization or a gauge invariant
projection method \cite{deser}. We also note that the one loop result
is  completely
independent of the presence or absence of a tree level Chern-Simons
term in $2+1$ dimensions and, therefore, holds even for pure
QED$_{3}$. Furthermore, let us note that while this has been an exact
result at one loop, it can also be easily checked within the hard thermal
loop approximation \cite{pisarski} where
\begin{equation}
\Pi_{\rm T\, (HTL)}^{(1)(\beta)} (0) \simeq - C(D)e^{2} \int
\frac{d^{D-1}k}{(2\pi)^{D-1}}\,
\frac{n_{F}(\omega_{k})}{\omega_{k}}\left((D-3) + \frac{\vec{k}^{2}
\vec{p}^{\, 2}}{(\vec{k}\cdot \vec{p}\, )^{2}}\right) = 0
\end{equation}
\begin{figure}
\centerline{\includegraphics[width = 14cm, height = 9cm]{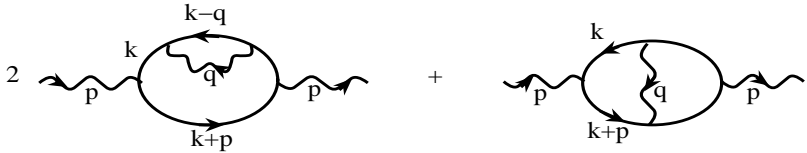}}
\vskip -4cm
\caption{Two loop diagrams for photon self-energy}
\end{figure}

We will show next that the magnetic mass also vanishes at two loops in
QED$_{3}$  with a  Chern-Simons term. The two
loop photon self-energy diagrams (see fig. 2) can be obtained from the
one  loop box
diagrams by connecting the photon lines in all possible ways. We note
that at tree level, the photon propagator, in a covariant gauge, has
the form
\begin{equation}
D_{\mu\nu}^{(0)} (p) = \frac{1}{p^{2}+\kappa^{2}} \left[(\delta_{\mu\nu} -
\frac{p_{\mu}p_{\nu}}{p^{2}}) - \kappa \epsilon_{\mu\nu\lambda}
\frac{p_{\lambda}}{p^{2}}\right] +
\xi\,\frac{p_{\mu}p_{\nu}}{(p^{2})^{2}}\label{gaugepropagator} 
\end{equation}
where $\xi$ is the gauge fixing parameter and $p_{0} = 2\pi nT =
\frac{2\pi n}{\beta}$. Furthermore, since the four photon amplitude
at one loop (sum of the box diagrams) is gauge invariant, it vanishes
when it is contracted with the momentum associated with a photon
line. Therefore, the only two terms from the photon propagator in
(\ref{gaugepropagator})  that
can contribute to the two loop self-energy are the $\delta_{\mu\nu}$
and the $\epsilon_{\mu\nu\lambda}$ terms. With these simplifications
in mind, we obtain from the two loop photon self-energy, in the static
limit,
\begin{equation}
\Pi_{\rm T}^{(2)} (0) =  4 e^{4} \int
\frac{d^{3}q}{(2\pi)^{3}} \frac{1}{q^{2} + \kappa^{2}} \int
\frac{d^{3}k}{(2\pi)^{3}}\,\frac{\partial}{\partial k_{i}}
\left[\frac{k_{i}}{(k^{2} + m^{2})^{2}} - \frac{k_{i}(q^{2} +
4m(\kappa-m))}{(k^{2}+m^{2})^{2}
((k+q)^{2}+m^{2})}\right] = 0
\end{equation}
This result holds for both the zero temperature as well as the finite
temperature parts and we note that this is true even when $\kappa = 0$
and,  therefore, the
vanishing of the magnetic mass, at two loops, holds even for pure
QED$_{3}$. This has to be contrasted with the behavior of the parity
violating part of the self-energy which has an infrared divergence
when $\kappa\rightarrow 0$ \cite{brandt}. Therefore, we see that
unlike the parity violating part of the self-energy, the parity
conserving part has a better infrared divergence behavior. It is worth
pointing out here that this result can also be obtained in a simple
manner using the Ward identities of the theory. Let us also
note that the Chern-Simons term does have a nontrivial contribution to
the  parity conserving part of the self-energy, $\Pi_{ij}^{\rm PC}$
\cite{van}. However, surprisingly, it does not contribute to the
magnetic mass (\ref{definition}).

\section{Vanishing of magnetic mass to all orders} 

Normally, an all orders proof of a result in a gauge theory is
simplified enormously through the use of Ward identities. However, at
finite temperature, the non-analyticity of the amplitudes at the
origin in the energy-momentum space leads to difficulties
\cite{das,weldon}.  For example,
let us consider the $N$-point photon amplitude which would satisfy a
relation of the form
\begin{equation}
p_{\alpha,\mu_{\alpha}} \Gamma_{\mu_{1},\cdots ,\mu_{N}} (p_{1},\cdots , p_{N})
= 0,\qquad \mu_{\alpha} = 0,1,2\label{ward}
\end{equation}
In the static limit where all the external energies vanish, on the
other hand, we can write the amplitude as $\Gamma_{(m,n)}$ with
$m+n=N$ where $m$ represents the number of time  indices while $n$
corresponds to the number of space indices. In this case, the
Ward identity (\ref{ward}) takes the form
\begin{equation}
p_{\alpha,i_{\alpha}} \Gamma_{(m,i_{1},\cdots ,i_{n})} (\cdots ,
p_{1},\cdots , p_{n}) = 0,\qquad i_{\alpha} = 1,2
\end{equation}
and one can formally argue that, for small
$p_{\alpha,i_{\alpha}}$, the amplitude behaves like \cite{ward,coleman}
\begin{equation}
\Gamma_{(m,i_{1},\cdots ,i_{n})} (\cdots ,p_{1},\cdots ,p_{n}) \sim
O(p_{1}\cdots p_{n})
\end{equation}
This behavior is, in fact, explicitly seen at one loop in the two point and the
four point functions for the photon in the static limit
\cite{brandt1}. In  the long
wave length limit where the spatial components of the momenta vanish,
on the other hand, we have much more limited information. Furthermore,
even if we know the behavior of the amplitudes in the static and the
long wave limits independently, this information is not very useful in
constructing higher loop amplitudes, where the energy and momenta of
the  internal photon lines are integrated over all possible values.

Therefore, we pursue the following strategy in proving, to all orders,
that the magnetic mass vanishes at finite temperature. First, let us
look at  the tree level propagator for the fermion
in (\ref{fermionpropagator}) and note that
\begin{equation}
\frac{\partial S^{(0)}(p)}{\partial p_{\mu}} = - S^{(0)}(p)
\gamma_{\mu} S^{(0)}(p)\label{basicidentity1}
\end{equation}
Namely, we see that, much like at zero temperature \cite{ward}, at finite
temperature, differentiating the tree level fermion propagator
is equivalent to introducing a tree level photon vertex with zero
energy-momentum 
(up to the coupling). This can also be rewritten in the more
familiar form
\begin{equation}
\frac{\partial (S^{(0)})^{-1} (p)}{\partial p_{\mu}} = 
\gamma_{\mu}\label{basicidentity2} 
\end{equation}
which says that, at the tree level, differentiating the fermion two
point function gives rise to a vertex with zero energy and  momentum, up to the
coupling constant.

\begin{figure}
\centerline{\includegraphics[width = 14cm, height = 9cm]{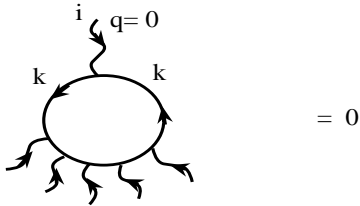}}
\vskip -4cm
\caption{Vanishing of an amplitude with one space like photon index
carrying zero energy and momentum}
\end{figure}

Let us next note that we can write the one loop $N$-point photon
amplitude as
\begin{equation}
\Gamma_{\mu_{1},\cdots ,\mu_{N}} (q_{1},\cdots ,q_{N}) = \int
\frac{d^{3}k}{(2\pi)^{3}}\,\tilde{\Gamma}_{\mu_{1},\cdots ,\mu_{N}}
(k;q_{1},\cdots ,q_{N}) = \frac{1}{\beta} \sum_{n} \int
\frac{d^{2}k}{(2\pi)^{2}}\,\tilde{\Gamma}_{\mu_{1},\cdots ,\mu_{N}}
(k;q_{1},\cdots ,q_{N})
\end{equation}
where, as mentioned earlier, $k_{0} = \frac{(2n+1)\pi}{\beta}$. Using
the relations in (\ref{basicidentity1})-(\ref{basicidentity2}), we see
that the amplitude with an additional external photon with a space index
and  carrying zero energy and momentum is obtained to be (see fig. 3)
\begin{equation}
\Gamma_{i,\mu_{1},\cdots ,\mu_{N}} (0,q_{1},\cdots ,q_{N}) =
- \frac{e}{\beta} \sum_{n} \int
\frac{d^{2}k}{(2\pi)^{2}}\,\frac{\partial}{\partial
k_{i}}\,\tilde{\Gamma}_{\mu_{1},\cdots ,\mu_{N}} (k;q_{1},\cdots ,q_{N}) =
0,\qquad N\geq 2
\end{equation}

This shows that an $N+1$ point amplitude with a 
photon line 
carrying  a space index and zero energy-momentum vanishes for $N\geq
2$, independent
of the values of the energy and momenta of the other photon lines. The
restriction, $N\geq 2$, comes from the ultraviolet convergence of the
integrand and, at one
loop, we have no infrared divergence problem since the fermion is
massive. We note that the odd point photon amplitudes
vanish by charge conjugation invariance (Furry's theorem) for any
value of the external energy and momentum, but 
this result shows that even point photon amplitudes also vanish when
one of the external photon lines has a space index and carries zero energy
and momentum. It follows now, from this as well as charge
conjugation invariance, that any fermion loop with an external photon
carrying a space index and zero energy-momentum gives a vanishing
contribution  in a
complicated diagram such as fig. 4, where the external photon is 
attached to an internal fermion line that is not a continuation of the
external fermion lines.  
\begin{figure}
\centerline{\includegraphics[width = 14cm, height = 9cm]{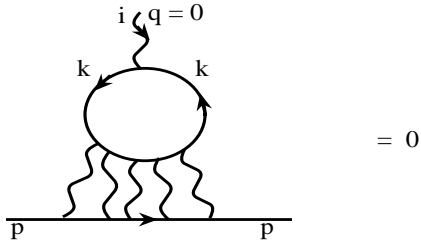}}
\vskip -4cm
\caption{Vanishing of a vertex diagram where the external photon with
a space index and 
zero energy-momentum is attached to an internal fermion line
that is not a continuation of the external fermion lines.}
\end{figure}

As a result, using (\ref{basicidentity1})-(\ref{basicidentity2}), it
can be  shown in a straightforward diagrammatic 
manner that, to all orders, we can write the three point
photon-fermion-fermion vertex with the photon carrying a space index
and  zero energy-momentum as
\begin{equation}
\Gamma_{i} (p,-p,0) = e\,\frac{\partial S^{-1} (p)}{\partial
p_{i}}\label{completeidentity}
\end{equation}
where $\Gamma_{i}$ and $S^{-1}(p)$ represent respectively the vertex
 and  the fermion self-energy to all orders. This is like the zero
temperature Ward identity \cite{ward}, but holds for a space index,
 which is  what we will need for our proof.

\begin{figure}
\vskip -2cm
\centerline{\includegraphics[width = 14cm, height = 9cm]{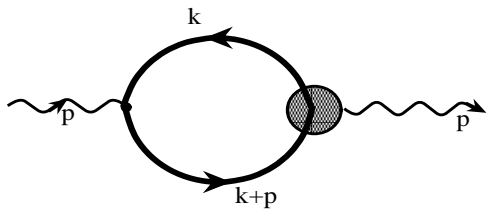}}
\vskip -4cm
\caption{Schwinger-Dyson relation for photon self-energy. The internal
heavy lines represent the full fermion propagator while the blob represents
the complete vertex.}
\end{figure}

With these, let us look at the photon self-energy to all orders given
by the Schwinger-Dyson equation, fig. 5, which leads to
\begin{eqnarray}
\Pi_{\rm T} (0) & = & \frac{e}{\beta} \sum_{n} \int
\frac{d^{2}k}{(2\pi)^{2}}\, {\rm tr}\,\gamma_{i}S(k)\Gamma_{i}(k,-k,0)
S(k)\nonumber\\
 & = & - \frac{e^{2}}{\beta} \sum_{n} \int
\frac{d^{2}k}{(2\pi)^{2}}\,\frac{\partial}{\partial k_{i}}\,({\rm
tr}\, \gamma_{i} S(k))
\end{eqnarray}
The finite temperature part of this integrand is well behaved and,
being a total divergence, vanishes, namely,
\begin{equation}
\Pi_{\rm T}^{(\beta)} (0) = 0
\end{equation}
to all orders. Since the loop involves massive fermions, there is no
problem of infrared divergence in this case. The temperature
independent part  of this expression has
a linear divergence which, as we have mentioned in the last section, can
be regularized to zero in $2+1$ dimensions. We note here that the
vanishing of the finite temperature part to all orders appears
naturally, in this description, to hold in any dimension.

\section{Summary}

In this paper, we have shown that, at finite temperature, the magnetic mass
vanishes at one loop in QED in any dimension. In $2+1$ dimensions, in
addition, the zero temperature part can be regularized to zero. In
$2+1$ dimension, this result is independent of the presence of a tree
level Chern-Simons term. We have calculated and shown explicitly that
this result also holds at two loops. We have given a simple proof to
show that the magnetic mass vanishes to all orders at finite
temperature  in this theory. This result also holds for QED in any dimension.

\newpage

\noindent{\bf Acknowledgment:}

One of us (AD) would like to thank Profs. J. Frenkel and S. Okubo for
many helpful discussions. This work was supported in part by US DOE
Grant number DE-FG 02-91ER40685 and by CAPES, Brasil.

\end{document}